\documentclass[sigconf]{acmart}

\newcommand{\squishlist}{
   \begin{list}{$\bullet$}
    { \setlength{\itemsep}{0pt}      \setlength{\parsep}{0pt}
      \setlength{\topsep}{-3pt}       \setlength{\partopsep}{0pt}
      \setlength{\listparindent}{-2pt}
      \setlength{\itemindent}{-5pt}
      \setlength{\leftmargin}{1em} \setlength{\labelwidth}{0em}
      \setlength{\labelsep}{0.5em} } }

\newcommand{\squishend}{
    \end{list}  }

\setcopyright{none}
\renewcommand\footnotetextcopyrightpermission[1]{}
\title{Multi-Agent Memory from a Computer Architecture Perspective: Visions and Challenges Ahead}

\author{Zhongming Yu}
\email{zhy025@ucsd.edu}
\affiliation{%
  \institution{University of California, San Diego}
  \city{San Diego}
  \state{California}
  \country{USA}}

\author{Naicheng Yu}
\email{n7yu@ucsd.edu}
\affiliation{%
  \institution{University of California, San Diego}
  \city{San Diego}
  \state{California}
  \country{USA}}

\author{Hejia Zhang}
\email{hez024@ucsd.edu}
\affiliation{%
  \institution{University of California, San Diego}
  \city{San Diego}
  \state{California}
  \country{USA}}

\author{Wentao Ni}
\email{w2ni@ucsd.edu}
\affiliation{%
  \institution{University of California, San Diego}
  \city{San Diego}
  \state{California}
  \country{USA}}

\author{Mingrui Yin}
\email{m3yin@ucsd.edu}
\affiliation{%
  \institution{University of California, San Diego}
  \city{San Diego}
  \state{California}
  \country{USA}}

\author{Jiaying Yang}
\email{jesyjy7@gatech.edu}
\affiliation{%
  \institution{Georgia Institute of Technology}
  \city{Atlanta}
  \state{Georgia}
  \country{USA}}

\author{Yujie Zhao}
\email{yuz285@ucsd.edu}
\affiliation{%
  \institution{University of California, San Diego}
  \city{San Diego}
  \state{California}
  \country{USA}}

\author{Jishen Zhao}
\email{jzhao@ucsd.edu}
\affiliation{%
  \institution{University of California, San Diego}
  \city{San Diego}
  \state{California}
  \country{USA}}

\acmConference[Architecture 2.0 '26]{Architecture 2.0 Workshop on AI for Computing Systems Design}{March 23, 2026}{Pittsburgh, PA, USA}

\begin{document}

\begin{abstract}
As LLM agents evolve into collaborative multi-agent systems, their memory requirements grow rapidly in complexity. This position paper frames multi-agent memory as a computer architecture problem. We distinguish shared and distributed memory paradigms, propose a three-layer memory hierarchy (I/O, cache, and memory), and identify two critical protocol gaps: cache sharing across agents and structured memory access control. We argue that the most pressing open challenge is multi-agent memory consistency. Our architectural framing provides a foundation for building reliable, scalable multi-agent systems.

\end{abstract}

\maketitle

\section{Introduction}
Large language model (LLM) agents~\cite{yao2022react,shinn2023reflexion} are quickly moving from ``single agent'' tools~\cite{schick2023toolformer}
to \textbf{multi-agent systems}~\cite{guo2024large, hong2023metagpt}: tool-using agents~\cite{wu2024autogen}, planner--orchestrator stacks~\cite{langgraph_overview},
debate teams~\cite{autogen_debate,chan2023chateval}, and specialized sub-agents that collaborate to solve tasks~\cite{langchain_subagents,schmidgall2025agent}. At
the same time, the \emph{context} these agents operate within is becoming more
complex: longer histories, multiple modalities, structured traces, and
customized environments. This combination creates a bottleneck that looks
surprisingly familiar to computer architects: \textbf{memory}.

In computer systems, performance and scalability are often limited not by
compute but by memory hierarchy, bandwidth, and consistency. Multi-agent systems
are heading toward the same wall---except their ``memory'' is not raw bytes, but
semantic context used for reasoning. This position paper frames multi-agent
memory as a computer architecture problem and highlights key protocol and
consistency gaps.

\begin{figure}[t]
  \centering
  \includegraphics[width=1.0\linewidth]{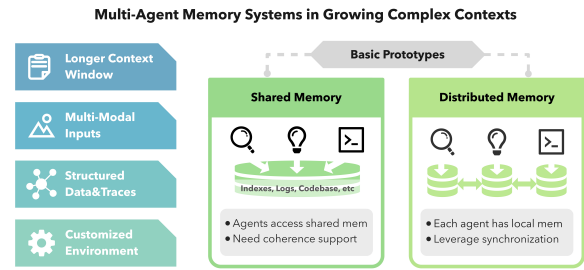}
  \caption{Two fundamental multi-agent memory architectures for managing growing context complexity: shared memory and distributed memory.}
  \Description{Overview graphic motivating multi-agent memory challenges.}
  \label{fig:motivation}
\end{figure}

\section{Why Memory Matters: Context Is Changing}
LLM evaluations show that ``real'' context ability involves more
than simple retrieval; it requires multi-hop tracing, aggregation, and sustained
reasoning as context length scales. Multimodal benchmarks add images, diagrams, and
videos. Structured tasks introduce executable traces and schemas. Interactive
environments make \emph{environment state + execution} part of the memory
problem. The result is not a static prompt but a dynamic, multi-format, partially
persistent memory system.

%\begin{itemize}
\squishlist
\item \textbf{Longer context windows:} Suites like RULER \cite{hsieh2024ruler}
  emphasize reasoning over long histories, not just retrieval.
\item \textbf{Multimodal inputs:} Benchmarks such as MMMU \cite{yue2024mmmu}
  and\\ VideoMME \cite{fu2025video} require joint reasoning over images and videos.
\item \textbf{Structured data \& traces:} Text-to-SQL datasets like Spider
  \cite{yu2018spider} and BIRD \cite{li2023can} show that agents increasingly
  operate over structured, executable traces.
\item \textbf{Customized environments:} Evaluations such as SWE-bench
  \cite{jimenez2023swe} and OSWorld \cite{xie2024osworld} stress long-horizon
  state tracking and grounded actions.
\squishend
%\end{itemize}

As such, context is no longer a static prompt; it is a dynamic
memory system with bandwidth, caching, and coherence constraints.

\section{Shared vs. Distributed Agent Memory}
Here we name two basic prototypes that mirror
classical memory systems. In \emph{shared memory}, all agents access a shared
pool (e.g., a shared vector store or document database). In \emph{distributed
memory}, each agent owns local memory and synchronizes selectively.

\textbf{Shared memory} makes knowledge reuse easy but requires coherence
support; without coordination, agents overwrite each other, read stale
information, or rely on inconsistent versions of shared facts. \textbf{Distributed
memory} improves isolation and scalability but requires explicit synchronization;
state divergence becomes common unless carefully managed. Most real systems sit
between these extremes: local working memory with selectively shared artifacts.

\begin{figure}[t]
  \centering
  \includegraphics[width=1.0\linewidth]{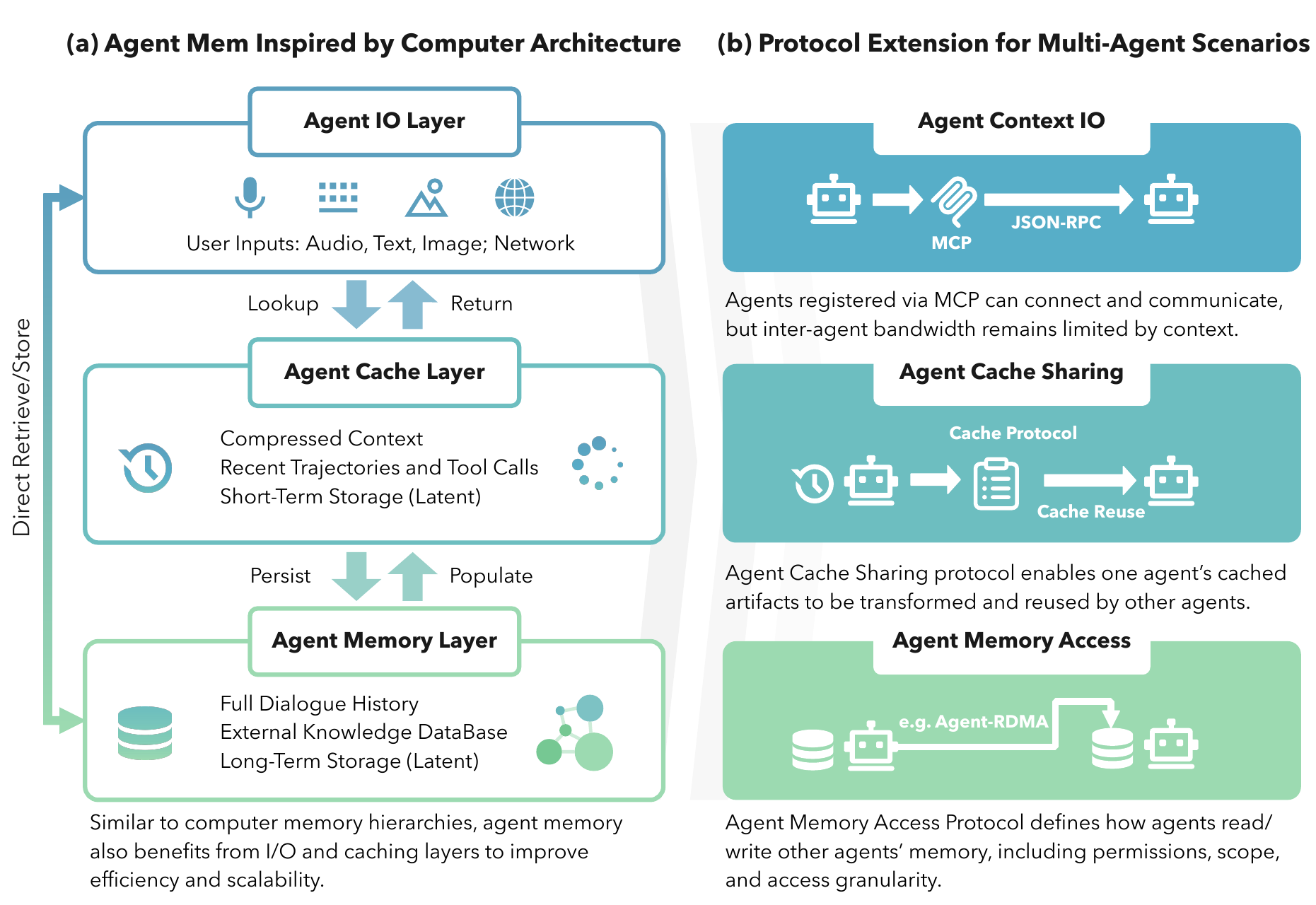}
  \caption{Agent memory hierarchy and protocol framing.}
  \Description{Diagram of agent I/O, cache, and memory layers with protocol links.}
  \label{fig:memprotocol}

\end{figure}

\vspace{-0.3em}
\section{An Architecture-Inspired Memory Hierarchy}
Computer architecture teaches a practical lesson: systems are not designed around ``one
memory.'' Instead, they are built as \textbf{memory hierarchies} with layers optimized for
latency, bandwidth, capacity, and persistence. A useful mapping for agents is as follows:

\textbf{Agent I/O layer:} Interfaces that ingest and emit information (audio,
text documents, images, network calls). \textbf{Agent cache layer:} fast,
limited-capacity memory for immediate reasoning (compressed context, recent
tool calls, short-term latent storage such as KV caches and embeddings).
\textbf{Agent memory layer:} large-capacity, slower memory optimized for
retrieval and persistence (full dialogue history, vector DBs, graph DBs, and
document stores).

This framing emphasizes a key principle: \textbf{agent performance is an
end-to-end data movement problem}. If relevant information is stuck in the wrong
layer (or never loaded), reasoning accuracy and efficiency degrade. As in
hardware, caching is not optional.

\vspace{-0.3em}

\section{Protocol Extensions for Multi-Agent Scenarios}
Architecture layers need \emph{protocols}. Many systems rely on connectivity
protocols, but inter-agent bandwidth remains limited by message passing. This
layer is best viewed as \emph{agent context I/O}, e.g. MCP~\cite{anthropic_mcp_intro}. That is
necessary---but not sufficient. Two missing pieces stand out.

\textbf{Missing piece 1: Agent cache sharing protocol.} Recent work explores KV
cache sharing~\cite{liu2024droidspeak, fu2025cache, ye2025kvcomm}, but we lack a principled protocol for sharing cached artifacts
across agents. The goal is to enable one agent's cached results to be
transformed and reused by another, analogous to cache transfers in
multiprocessors.

\textbf{Missing piece 2: Agent memory access protocol.} Agentic memory frameworks~\cite{packer2023memgpt, xu2025mem, chhikara2025mem0, qian2024memorag} propose many strategies for maintaining and optimizing LLM agents' memory. Yet even when some frameworks support
shared state, the standard access protocol (permissions, scope, granularity) remains
under-specified. Key questions include: Can one agent read another's long-term
memory? Is access read-only or read-write? What is the unit of access: a
document, a chunk, a key-value record, or a trace segment?

\begin{figure}[t]
  \centering
  \includegraphics[width=0.98\linewidth]{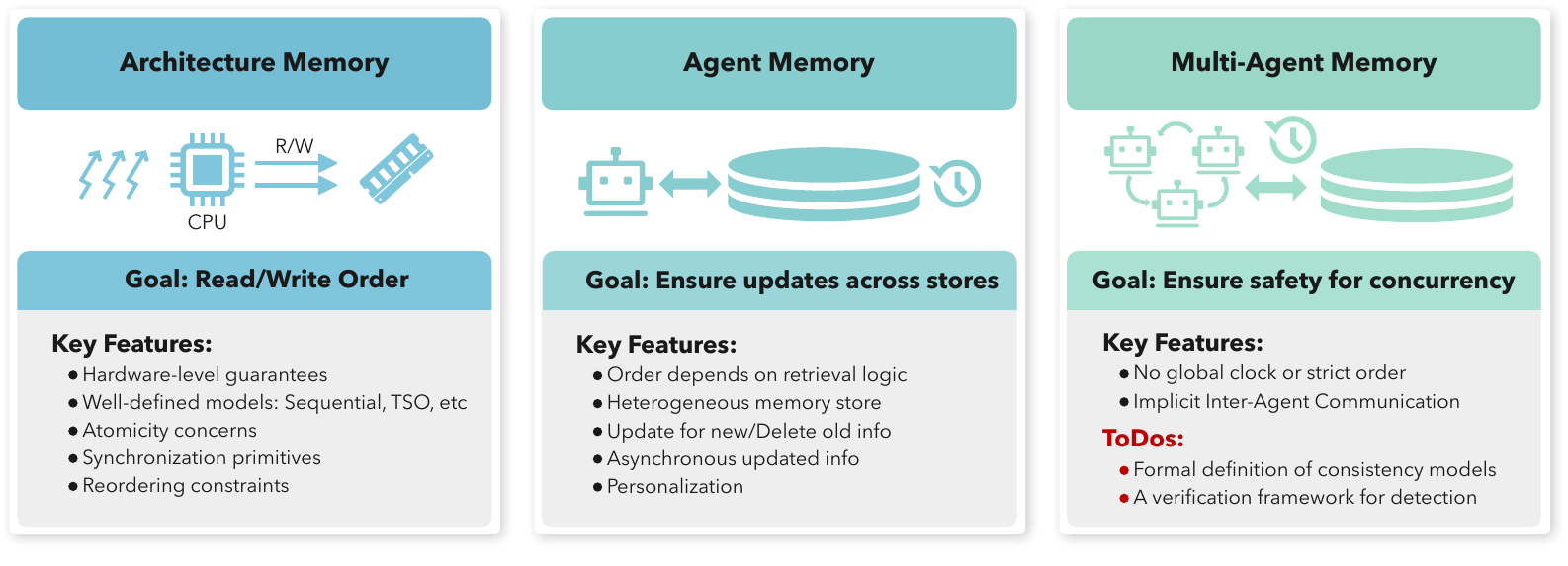}
  \caption{ Consistency model comparison from traditional memory architecture to multi-agent memory.}
  \Description{Comparison chart of memory consistency models.}
  \label{fig:consistency}
\end{figure}

\vspace{-0.3em}

\section{The Next Frontier: Multi-Agent Consistency}

In computer architecture, \emph{memory consistency models}~\cite{sorin2011primer}
specify the order in which memory operations issued by one processor become
visible to others—governing not \emph{what} value a read returns, but
\emph{when} and \emph{in what order} concurrent writes may be observed.
Decades of systems research have produced a precise hierarchy of such
models (SC, TSO, Release Consistency) together with formal verification
techniques. Agent memory systems face an analogous challenge, yet no
equivalent formalism exists.

For a \emph{single} LLM agent, consistency requires that updates to
memory—across working, episodic, and semantic stores—propagate in a
causally order. Prior work on generative
agents~\cite{park2023generative} acknowledges the need for temporally
coherent retrieval, but does not formalize the ordering guarantees under
which retrieval is correct.

The problem compounds in \emph{multi-agent} settings, where multiple
agents concurrently read from and write to shared memory. Here,
consistency decomposes into two requirements. \textbf{Update-time
visibility and ordering} determines when an agent's writes become
observable to others, and in what order concurrent writes from different
agents may be observed—directly analogous to the ordering contracts of
hardware consistency models. \textbf{Read-time conflict resolution}
governs how an agent should reconcile conflicting or stale artifacts that
arise when records evolve across versions under concurrent revision.
Both requirements are harder than their hardware counterparts: memory
artifacts are semantically heterogeneous (evidence, tool traces, plans),
inter-agent dependencies are implicit rather than declared, and conflicts
are often semantic and coupled to environment state.

Despite this complexity, \emph{multi-agent memory consistency has not
been formally defined}, and no framework exists to detect or classify
consistency violations in practice. 

% \textbf{Why agent consistency is harder:} state is not a scalar value but a
% plan, a summary, a retrieval result, or a tool trace. Writes may be speculative
% or wrong. Conflicts are semantic contradictions, and freshness depends on
% environment state. A practical direction is to define consistency around the
% artifacts agents actually share---cached evidence, tool traces, plans, and
% long-term records---across both shared and distributed memory setups. The layer
% should expose consistency models (session, causal, eventual semantic, and
% stronger guarantees for committed outputs), communication primitives beyond
% message passing, and conflict-resolution policies.

\vspace{-0.3em}

\section{Conclusion}
Many agent memory systems today resemble human memory: informal, redundant, and
hard to control. To move from ad-hoc prompting to reliable multi-agent systems,
we need better hierarchies, explicit protocols for cache sharing and memory
access, and principled consistency models that keep shared context coherent.
We believe this architecture framing is a foundational research direction for
next-generation agent systems.

\newpage

\bibliographystyle{ACM-Reference-Format}

\bibliography{ref}

\end{document}